# Dark exciton energy splitting in monolayer WSe$_2$: insights from time-dependent density-functional theory

Jia Shi,[1,2] Volodymyr Turkowski,[2*] Duy Le,[2] Talat S. Rahman[2]

[1] *Department of Physics, University of Science and Technology Beijing, Beijing, 100083 China*

[2] *Department of Physics, University of Central Florida, Orlando, Florida, 32816 USA*

We present here a formalism based on time-dependent density-functional theory (TDDFT) to describe characteristics of both intra- and inter-valley excitons in semiconductors, the latter of which had remained a challenge. Through the usage of an appropriate exchange-correlation kernel (nanoquanta), we trace the energy difference between the intra- and inter-valley dark excitons in monolayer (1L) WSe$_2$ to the domination of the exchange part in the exchange-correlation energies of these states. Furthermore, our calculated transition contribution maps establish the momentum resolved weights of the electron-hole excitations in both bright and dark excitons thereby providing a comprehensive understanding of excitonic properties of 1L WSe$_2$. We find that the states consist of hybridized excitations around the corresponding valleys which leads to brightening of the dark excitons, i.e., significantly decreasing their lifetime which is reflected in the PL spectrum. Using many-body perturbation theory, we calculate the phonon contribution to the energy bandgap and the linewidths of the excited electrons, holes and (bright) exciton to find that as the temperature increases the bandgap significantly decreases, while the linewidths increase. Our work paves for describing the ultrafast charge dynamics of transition metal dichalcogenide within an *ab initio* framework.

*Corresponding author: Volodymyr.Turkowski@ucf.edu

## 1. Introduction

Monolayer transition metal dichalcogenides (TMDs) are a class of two-dimensional (2D) materials with unique, most notably optical, properties that have drawn considerable interest in recent years.[1,2] Compared to the case of the bulk, monolayer TMDs exhibit a direct band gap at K points in the Brillouin zone with highly efficient luminescence.[3,4] The systems also demonstrate high electron mobility and room-temperature current on/off ratio, with potential field-effect transistor applications.[5] As one of most actively studied TMDs, it is known that 1L $WSe_2$, a three-atom thick system, displaying a pronounced absorption region[6] and a direct band gap at two time-reversal valleys (K, K′) in the Brillouin zone.[7-10] Due to spin-orbit coupling (SOC), the bottom of the conduction and top of the valence band are each split into two sub bands with opposite spins. The valence band splitting is wide, in the hundred meV range, while that of the bottom of conduction band is in the tenths of meV.[9] The splitting of conduction band, in particular, creates an ideal environment for forming intra-valley bright and intra- and inter-valley dark excitons (shown in Figure 1). The properties of these excitons can be tuned by changing valley polarization and by the interaction between the different (orbital, spin, valley) degrees of freedom,[11,12] giving rise to exotic effects such as valley optical selection rules[13]. The intravalley bright (opposite spin for electron and hole) or dark (same spin for electron and hole) excitons consist of the electron-hole pairs from the same K valley, constrained by the spin orientation, with the bright exciton giving the major contribution to the photoluminescence (PL). On the other hand, the intervalley dark (opposite spin for electron and hole) exciton is composed of electrons and holes from K and K′ valleys.[12-14] Because the recombination of the electron and hole from different valleys is forbidden by momentum conservation law, the emission from (phonon-assisted) recombination of the intervalley dark exciton is very low. In other words, the lifetime of dark excitons is much longer than that of the bright ones.[15,16]
Gate- and magnetic-field based measurements show PL peaks from both bright and dark excitons, with almost measurement-independent, 16meV, difference in energy for the intra- and inter-valley dark excitons.[17] In a combined experimental+theoretical study using a semiquantitative approach in which wavefunctions and energies from DFT were used as input for many-body analysis, Liu et al.[18] showed that the binding energy of the intervalley dark exciton is 10 meV smaller than that of the intravalley dark exciton because of the difference in the Coulomb exchange energy as a result of different orientations of the electron and hole in these two systems (for intra- and intervalley exciton state transitions, see work [19]). Similarly, Yang et al. found experimentally that intervalley exciton has smaller binding energy by ~ 9 meV.[20] The dark excitonic states are seen in the PL spectrum, and the reason for the rather high signatures of the intra- and inter-valley dark exciton states in the PL despite a low recombination probability, is their relatively large population.[20]

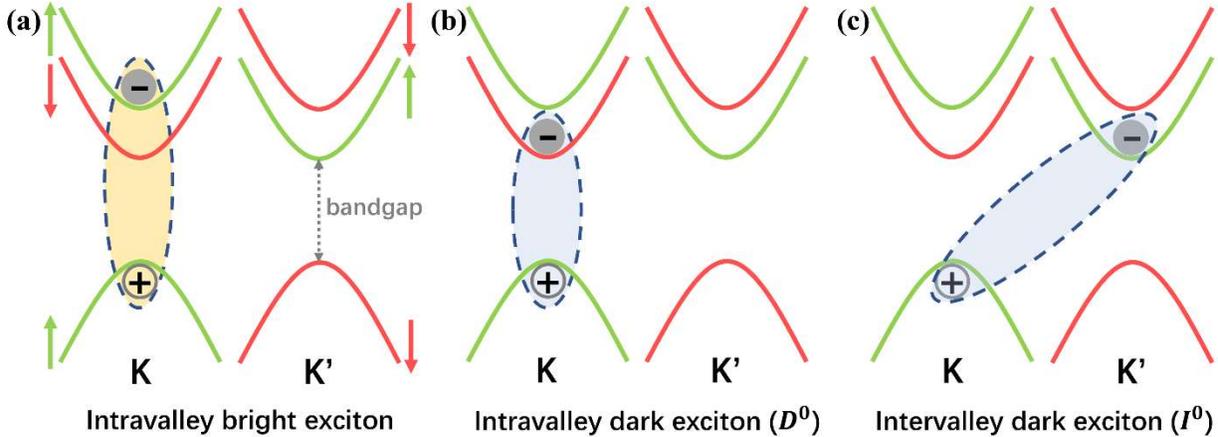

**Figure 1.** (Color online) Schematic representation of the (a) intravalley bright exciton, (b) intravalley dark exciton, and (c) intervalley exciton in 1L WSe$_2$. The band in red and green represent the opposite spin polarizations (depicted by the arrows) at time-reversal K and K′ points in BZ.

Available first-principles studies of excitons and the optical spectra in 1L WSe$_2$, mostly based on the GW-Bethe-Salpeter equation (GW-BSE) approach,[21-23] find the binding energy of the direct (bright) exciton to be in reasonable agreement with experiments.[24] Despite this remarkable achievement, the GW-BSE methodology is more suitable to calculations of direct (intravalley) excitons. Furthermore, the optical absorption of extended systems studied by the many-body BSE theory is computationally demanding, requiring calculations for an enormous number of k-points to solve the two-particle Green's functions.[25] There is thus an urgent and economic need in theoretical *ab initio* efforts for developing computationally efficient techniques for reliable calculations of both indirect (intervalley) dark excitons and the usual bright excitons. As we document below, time-dependent density functional theory (TDDFT) is one such scheme.

As in TDDFT one solves an effective one-particle problem, it is technically much simpler than methods such as BSE. It is a theory of one-particle density that depends on one space vector and one time variable. Thus, the solution can be obtained in a feasible manner, provided an appropriate exchange-correlation (XC) potential is available. Formally, the linear-response TDDFT used for calculation of excited states incorporates many-body effects into the dynamical exchange-correlation kernel $f_{xc}(\mathbf{r}, t, \mathbf{r}'; t') = \delta v_{xc}(\mathbf{r}, t)/\delta n(\mathbf{r}', t')$, where $v_{xc}(\mathbf{r}, t)$ denotes the corresponding time-dependent exchange-correlation potential and $n(\mathbf{r}', t)$ is the charge density. The standard approximations for $v_{xc}$ (local density or generalized gradient form) typically predict the ground-states properties of extended system quite reliably, but fail to describe excitonic effects in the optical and energy-loss spectra.[26] Moreover, $f_{xc}$ is often defined either by empirical parameters (such as the long-range (LR) kernel)[27] or by some physically-motivated but not quite universal approximations (e.g., the exact-exchange kernel)[28]. What is needed is an "exact" (parameter-free) and technically simple $f_{xc}$ to accurately study excitonic and multi-excitonic effects in extended systems, such as two dimensional TMDs. From this point of view, the nanoquanta kernel derived from the BSE approach has an acceptable computational cost, as it employs the one-loop approximation for the susceptibilities, and has proven to be suitable for selected systems.[29-31]

In this work, we focus on describing the fundamental properties of intra and intervalley bright and dark excitons by using the nanoquanta kernel. We solve the TDDFT problem in the density-matrix representation and obtain results for the excitonic binding energies for the intra-valley bright and intra and intervalley dark excitons in 1L $WSe_2$. We also calculate the transition contribution maps to find the composition of these excitons in terms of the free electron-hole excitations. Finally, by adding phonon effects into the theory in the form of many-body perturbation theory we calculate the temperature-dependence of the electron and hole excitation energies and lifetimes of electrons, holes and excitons.

## 2. Methodology

2.1 Theory

*TDDFT for intervalley excitons*

The Casida equation formalism for TDDFT allows one to calculate the eigenenergies and eigenstates of the system by solving the corresponding eigenproblem.[32] However, so far a TDDFT formalism for describing non-conserving-momenta transitions, including inter-valley excitons, is absent. To derive TDDFT equation for the intervalley bound states, we begin with the time-dependent Kohn-Sham equation[33-35]:

$$i\frac{\partial \Psi_k^v(r,t)}{\partial t} = H(r,t)\Psi_k^v(r,t), \tag{1}$$

where $k$ is the wave vector and the system Hamiltonian is given by

$$H[n](r,t) = -\frac{\nabla^2}{2m} + V_H[n](r,t) + V_{XC}[n](r,t) + erE(t), \tag{2}$$

which includes the kinetic (first), Hartree (second), and XC (third) potential terms, as well as external electric field part (the last term). The self-consistent electron density is used to solve Eq. (1):

$$n(r,t) = \sum_{l,|k|<k_F}|\Psi_k^l(r,t)|^2, \tag{3}$$

where $l$ is combined band and spin index.

To derive the exciton eigenenergy equation it is convenient to use the density-matrix formalism[33-35] in which one expands $\Psi_k^v(r,t)$ in the basis of the wavefunctions $\psi_k^l(r)$ obtained from DFT:

$$\Psi_k^v(r,t) = \sum_{l=v,c} c_k^l(t)\psi_k^l, \tag{4}$$

where $c_k^l$ are time-dependent coefficients. From Eqs. (1) and (4) we thus obtain:

$$i\frac{\partial c_k^m}{\partial t} = \sum_{l=v,c;q} H_{kq}^{ml} c_q^l, \tag{5}$$

where

$$H_{kq}^{ml}(t) = \int \psi_k^{m*}(r) H[n](r,t) \psi_q^l(r) dr \tag{6}$$

are matrix elements of the Hamiltonian.

Instead of solving Eq. (5), it is convenient to solve the equation for the density matrix

$$\rho_{kq}^{lm}(t) = c_k^l(t) c_q^{m*}(t), \tag{7}$$

that contains more physical information such as the state occupancies which are given by its diagonal elements, while electron transitions (polarizations, including excitonic states) are given by the non-diagonal elements. The density matrix satisfies the Liouville equation

$$i\frac{\partial \rho_{kq}^{lm}(t)}{\partial t} = [H,\rho]_{kq}^{ml}(t) \equiv \sum_{n,p}\left(H_{kp}^{ln}(t)\rho_{pq}^{nm}(t) - \rho_{kp}^{ln}(t)H_{pq}^{nm}(t)\right). \tag{8}$$

The four types of the matrix elements (for valence $v$ and conduction $c$ bands): $\rho_{kk}^{vv}(t)$, $\rho_{kq}^{cv}(t)$, $\rho_{kq}^{vc}(t)$ and $\rho_{kk}^{cc}(t)$ are not all independent, since by definition:

$$1 = \int |\Psi_k^l(r,t)|^2 = \rho_{kk}^{vv}(t) + \rho_{kk}^{cc}(t), \text{ and } \rho_{kq}^{cv}(t) = \rho_{qk}^{vc}(t). \tag{9}$$

We choose $\rho_{kq}^{cv}(t)$ and $\rho_{kk}^{cc}(t)$ as the two independent matrix elements, the equations of motion for which are

$$i\frac{\partial \rho_{kq}^{cv}(t)}{\partial t} = H_{kp}^{cc}(t)\rho_{pq}^{cv}(t) - \rho_{kp}^{cv}(t)H_{pq}^{vv}(t) + \left(1 - \rho_{qq}^{cc}(t) - \rho_{kk}^{cc}(t)\right)H_{kq}^{cv}(t), \quad (10)$$

$$\frac{\partial \rho_{kk}^{cc}(t)}{\partial t} = -2Im\left(H_{pk}^{vc}(t)\rho_{kp}^{cv}(t)\right). \quad (11)$$

The above equations describe dynamics of the system.

Since we are interested in the excitonic eigenstates lets us derive the eigenequation by linearizing the polarization equation (10):

$$i\frac{\partial \rho_{kq}^{cv}(t)}{\partial t} = (\varepsilon_k^c - \varepsilon_q^v)\rho_{kq}^{cv}(t) + \sum_{q<k_F,a,b}\int_0^t F_{kpqq}^{cvcv}(t,t')\rho_{qq}^{cv}(t')dt'$$

$$+ \sum_{q<k_F,a,b}\int_0^t F_{kpqq}^{cvcv}(t,t')\rho_{qq}^{vc}(t')dt' + \vec{d}_{kq}^{cv}\vec{E}(t), \quad (12)$$

where the function

$$F_{kpqq}^{lmab}(t,t') = \iint drdr'\,\psi_k^{m*}(r)\psi_p^l(r)\left(\frac{1}{|r-r'|}\delta(t-t') + f_{xc}(r,r',t,t')\right)\psi_q^{a*}(r')\psi_q^b(r') \quad (13)$$

describes scattering of two charges from states with momenta $k, p$ to those with momenta $q$ and $q$. These elements are defined by the XC kernel $f_{xc}(r,r',t,t')$. In Eq.(12), $\varepsilon_k^c$ and $\varepsilon_q^v$ are the spectra of the conduction and valence bands obtained from DFT. The excitonic states are described by the elements $\rho_{kp}^{cv}(t)$. In the spin-dependent case, for the spin triplet excitations the products of two functions $\psi_k^{m*}(r)\psi_q^b(r')$ and $\psi_p^l(r)\psi_q^{a*}(r')$ cossresponding to such transitions (like in the case of intervalley dark excitons, see below) in Eq. (13) have to be substituted by an antisymmetric combinations, e.g. $\psi_k^{m*}(r)\psi_q^b(r') \to \frac{1}{\sqrt{2}}\left(\psi_k^{m*}(r)\psi_q^b(r') - \psi_k^{m*}(r')\psi_q^b(r)\right)$, and similar for the second pair. In the Tamm-Dancoff approximation, $\rho_{qq}^{vc}(t)$ (de-excitation) terms can be neglected and the resulting Casida equation becomes:

$$(\omega - \varepsilon_k^c + \varepsilon_q^v)\rho_{kq}^{cv}(\omega) - \sum_{q<k_F} F_{kpqq}^{cvcv}(\omega)\rho_{qq}^{cv}(\omega) - \vec{d}_{kq}^{cv}\vec{E}(\omega) = 0. \quad (14)$$

Solution of the above equation yields the eigenenergies of the excited system, including exciton binding energies, when $\omega < \Delta_g$ ($\Delta_g$ is bandgap). Equation (14) is quite general, since it can describe intra- ($k = q$) and inter-valley ($k \neq q$) transitions. This is a generalization of the density-matrix TDDFT equation for excitons [34] for the case when the electron and hole have different momenta. Equation for the standard intra-valley excitons has been analyzed in previous works.[33-35]

We note that equation (14) for the inter-valley excitons can be solved separately for the diagonal and non-diagonal elements. Thus, one needs to find first the solution for the diagonal part $\tilde{\rho}_{qq}^{cv}(\omega)$ that satisfies the equation

$$(\omega - \varepsilon_k^c + \varepsilon_q^v)\tilde{\rho}_{kk}^{cv}(\omega) - \sum_{q<k_F} F_{kkqq}^{cvcv}(\omega)\tilde{\rho}_{qq}^{cv}(\omega) - \vec{d}_{kk}^{cv}\vec{E}(\omega) = 0. \tag{15}$$

The solution of the above equation is

$$\tilde{\rho}_{qq}^{cv}(\omega) = \sum_{q<k_F} B_{kq}^{-1} \vec{d}_{kk}^{cv}\vec{E}(\omega), \tag{16}$$

where

$$B_{kq}(\omega) = (\omega - \varepsilon_k^c + \varepsilon_q^v + i\delta)\delta_{kq} - F_{kkqq}^{cvcv}(\omega) \tag{17}$$

and $i\delta$ is a small imaginary part.

Substitution of equation (16) into equation (14) gives the equation for the inter-valley polarization $\rho_{kq}^{cv}(\omega)$.

Next, one can use the solution of the resulting equation to analyze the peaks in the absorption spectrum $A(\omega) \sim -\frac{1}{|\vec{E}(\omega)|} Im\rho_{kq}^{cv}(\omega)$ to find the intervalley exciton peak that corresponds to the pole of

$$\frac{1}{(\omega - \varepsilon_k^c + \varepsilon_q^v + i\delta)} \sum_{k,q<k_F} F_{kpqq}^{cvcv}(\omega) B_{kq}^{-1}(\omega) \vec{d}_{kk}^{cv}, \tag{18}$$

i.e., a pole of $\rho_{kq}^{cv}(\omega)$ below the gap which gives rise to bound states.

*XC kernels*

An important aspect of the above derivation is the XC kernel. In this work, we consider three different XC kernels:

1. Local kernel[36]: this is simply the phenomenological contact interaction

$$f_{XC}^{local}(\mathbf{r},\mathbf{r}') = -\alpha\delta(\mathbf{r}-\mathbf{r}'), \tag{19}$$

    where $\alpha$ is a parameter describing the strength of the TDDFT local electron-hole attraction

2. Long range (LR) kernel[36]: it is given by

$$f_{XC}^{LR}(r,r') = -\frac{1}{\varepsilon}\frac{1}{|r-r'|}, \tag{20}$$

where ε is the effective screening parameter of the electron-hole interaction.

3. Nanoquanta XC kernel derived from the BSE with the matrix elements (in momentum space) defined by [27]

$$f_{XCGG'}(\mathbf{q}) = \sum_{v_1,v_3,c_2,c_4,k_1,k_3} \frac{1}{f_{v_1k_1}-f_{c_2k_2}} \Phi^{-1}(v_1\mathbf{k}_1, c_2\mathbf{k}_2; \mathbf{G}) \mathcal{F}_{\substack{v_1c_1;v_2c_2 \\ k_1k_2;k_3k_4}} \Phi^{-1}(v_3\mathbf{k}_3, c_4\mathbf{k}_4; G'), \tag{21}$$

where $\mathbf{q} = \mathbf{k}_2 - \mathbf{k}_1 = \mathbf{k}_4 - \mathbf{k}_3$ and $\mathbf{G}$ are reciprocal vectors. Functions Φ in Eq. (21) are Fourier transforms of $\Phi(v_1\mathbf{k}_1, c_2\mathbf{k}_2; r) = \psi_{k_1}^{v_1}(r)\psi_{k_2}^{c_2*}(r)$ and functions $\mathcal{F}_{\substack{v_1c_1;v_2c_2 \\ k_1k_2;k_3k_4}}$ are

$$\mathcal{F}_{\substack{v_1c_1;v_2c_2 \\ k_1k_2;k_3k_4}} = -\iint drdr' \Phi(v_1\mathbf{k}_1, v_3\mathbf{k}_3; r) W(r,r') \Phi(c_2\mathbf{k}_2, c_4\mathbf{k}_4; r')$$

$$= -\frac{4\pi}{V} \sum_{G,G';k,k'} \frac{\epsilon_{GG'}^{-1}(\mathbf{q})}{|\mathbf{q}+G|^2} \langle c_2\mathbf{k}_2|e^{i(q+G)r}|c_4\mathbf{k}_4\rangle \langle v_3\mathbf{k}_3|e^{-i(q+G')r'}|v_1\mathbf{k}_1\rangle, \tag{22}$$

where $W(r,r')$ is screened Coulomb interactions, $V$ denotes the crystal volume and the inverse macroscopic dielectric function $\epsilon_{GG'}^{-1}(q)$ is approximated by using the one-loop charge susceptibilities.

*Effect of electron-phonon interaction*

As is the case with DFT calculations of system energetics, the above derivations are valid at very low (zero) temperature. Here, we include effects of the ambient temperature through electron-phonon and exciton-phonon interactions which provide temperature dependencies of the bandgap and the excitonic line width (inverse lifetime). Using second-order perturbation theory to include electron-phonon interaction in the calculated electronic band energy $E_k^l(T)$[37] leads to

$$E_k^l(T) = \varepsilon_k^l + \Sigma_k^{Fan,l}(\omega, T) + \Sigma_k^{DW,l}(T) \tag{23}$$

where

$$\Sigma_k^{Fan,l}(\omega, T) = \sum_{l',q,v} \frac{\left|g_{qv}^{(1)}(\mathbf{k}, l, l')\right|^2}{N_q} \left[\frac{N_{qv}(T) + 1 - f_{l',k-q}}{\omega - \varepsilon_{k-q}^{l'} - \omega_{qv} + i\delta} + \frac{N_{qv}(T) + f_{l',k-q}}{\omega - \varepsilon_{k-q}^{l'} + \omega_{qv} + i\delta}\right] \tag{24}$$

is the Fan part of the electron self-energy, and

$$\Sigma_k^{DW,l}(\omega, T) = \frac{1}{N_q} \sum_{q,v} g_{qv,-qv}^{(2)}(\mathbf{k}, l, l)(2N_{qv}(T) + 1) \tag{25}$$

is the Debye-Weller part of self-energy. In Eqs. (24) and (25), $N$ and $f$ are Bose and Fermi distribution functions, and $g_{qv}^{(1)}(k,l,l')$ and $g_{qv,q'v'}^{(2)}(k,l,l')$ are the first- and the second-order process scattering probabilities, where $q$ is phonon momentum and $v$ is phonon band. For example, in the second-order case the probability corresponds to the scattering process $|lk\rangle \to |l'k - q - q'\rangle \otimes |qv\rangle \otimes |q'v'\rangle$, i.e. electron state transforms into another electron state with two absorbed (emitted) phonons (for details, see Ref. [37]).

The shift of the band gap at the **K**-point obtained by subtracting $\varepsilon_k^c - \varepsilon_k^v$ from $E_k^c(T) - E_k^v(T)$, is given by:

$$\Delta = \Sigma_K^{Fan,c}(E_K^c, T) + \Sigma_K^{DW,c}(T) - \Sigma_K^{Fan,v}(E_K^v, T) - \Sigma_K^{DW,v}(E_K^c, T). \quad (26)$$

The phonon-induced exciton line broadening for state $l$ obtained using second order perturbation theory in exciton-phonon coupling is:

$$\gamma = \frac{1}{\pi} \sum_{q',v,\pm} \left| g_{qv}^{(1)}(K,l,l) \right|^2 \left( \frac{1}{2} \pm \frac{1}{2} + N_{\omega_{v,\pm q'}} \right) \delta\left(\hbar\omega - E^l(K+q') \mp \omega_{v,\pm q'}\right). \quad (27)$$

For further details of calculations of the effect of phonons on optical properties using semi-empirical methods see Refs. [38,39] and for first-principles implementation see Refs. [40-42].

2.2 Computational details

The DFT Kohn-Sham wavefunctions and eigenenergies were generated by the code QUANTUM ESPRESSO[43] and served as the input for the TDDFT code BEE (Binding Energy of Excitons)[35] developed in our group to obtain the associated fundamental parameters and solutions of the exciton eigenenergy equation. The Perdew-Burke-Ernzerhof (PBE) XC potential within the generalized gradient approximation (GGA)[44] was used in the DFT calculations. We used the norm-conserving pseudopotential to avoid inappropriate normalization of pseudo-wave functions.[45] A cutoff energy of 70 Ry and a $(12 \times 12 \times 1)$ k-point mesh in the 2D Brillouin zone (BZ) were used to optimize the crystal structure.[46] The 1L WSe$_2$ is modeled using a cell with (1×1) periodicity added to the vacuum slab ~12 Å along the c direction (shown in Figure 2). The optimized lattice constant $a$ was found to be 3.357 Å. Further, the band structure calculations in the presence of SOC were performed with fully relativistic pseudopotentials.[47] The calculations to generate the wavefunction with SOC were performed with (18 × 18 × 1) k-point grid in the BZ. Density functional perturbation theory (DFPT)[48] was used to calculate the dispersion of phonons and the electron-phonon scattering probabilities across the BZ using QUANTUM ESPRESSO. The phonon frequency calculations were converged for residual force less than 0.00001 eV/ Å$^{-1}$.

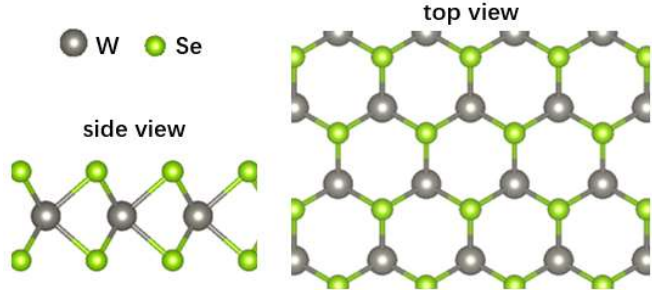

**Figure 2.** (Color online) Atomic structure of 1L WSe$_2$.

## 3. Results and discussions

3.1 Band structure and exciton binding energy

The calculated band structure of 1L WSe$_2$ is shown in Figure 3. Although the PBE potential underestimates the bandgap of 1L WSe$_2$ (1.46 eV which is less than the experimental value of 2.02eV,[49] our result regarding splitting of the bottom of the conduction band (~38.6 meV) owing to SOC are in good agreement with experimental observations and theoretical predictions.[9,50] Functionals that account for screening effects, such as the Heyd-Scuseria-Ernzerhof (HSE) hybrid functional, provide bandgap that agrees with the experimental data,[51] but with a high computational cost. Since the calculated binding energies (of the exciton) with respect to the bottom of sub-bands should not be significantly sensitive to the value of the bandgap, we have stayed with the PBE functional for our DFT calculations.

For further computational feasibility, we use a three-band approximation, which includes the top of valence band (V1) and two sub-bands of bottom of conduction band (C1, C2) split by the SOC effect, in the TDDFT calculation for the binding energy of excitons. Note that TDDFT results for the exciton binding energies are less sensitive to the number of used bands as compared to the BSE approach.[52] Figure 3 also displays the momentum-resolution of the three bands across the BZ. The highest level of the top of the valence band is located at K and K′ points, similarly the lowest levels of the conduction bands are also at these points, i.e. the system is a direct-gap semiconductor.

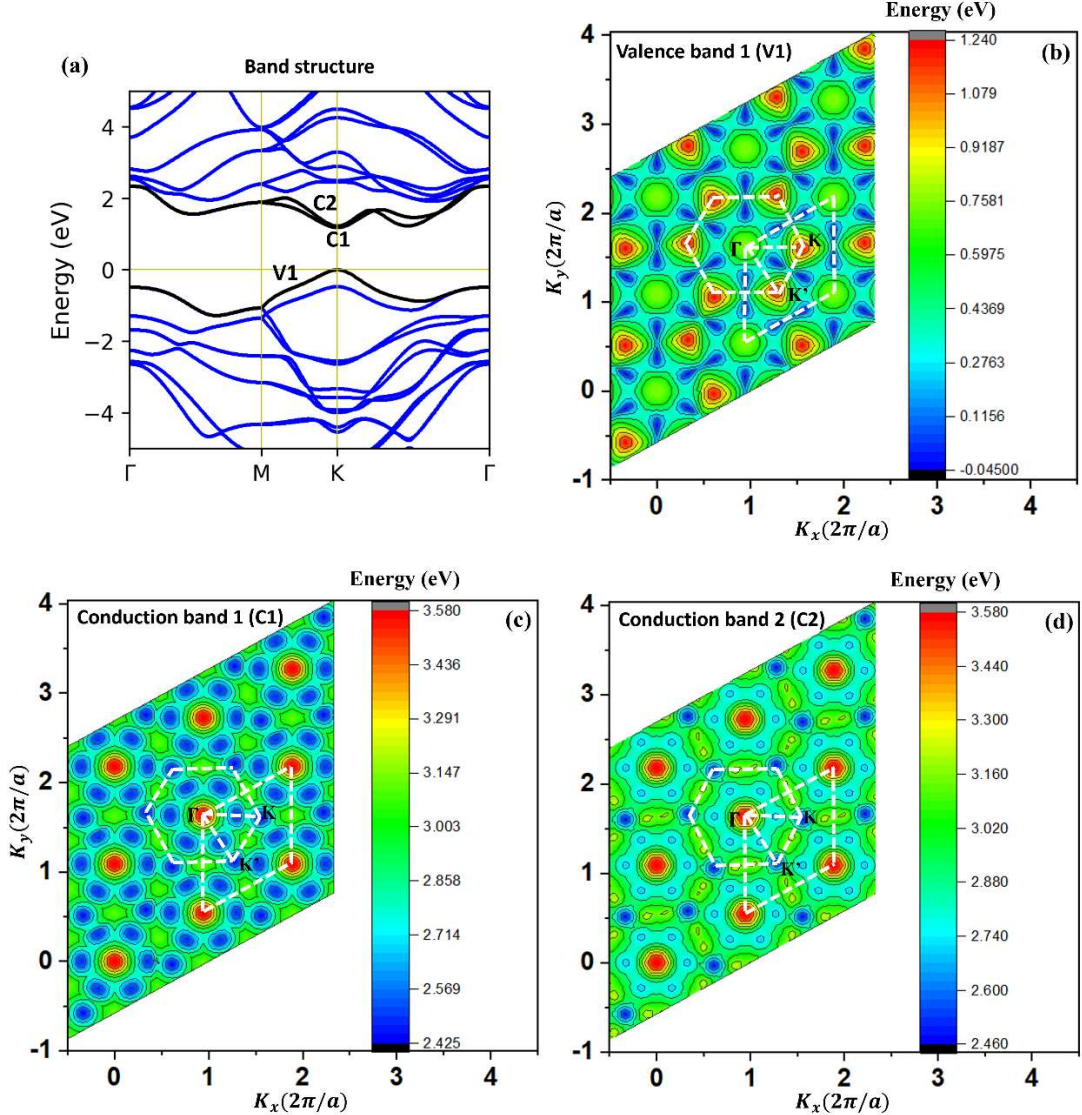

**Figure 3.** (Color online) The band structure of 1L WSe$_2$ along the path of the high-symmetry points (a). The three-band approximation used for TDDFT (black solids in (a)) in our calculations, which include the valence band (V1) (b), and the split conduction band (C1) (c) and conduction band (C2) (d). The full and the first Brillouin zones are shown in white,

The theoretical values of the exciton binding energy in 1L WSe$_2$ still lack trustworthy consistency in the community. The DFT-based method[53] and GW-BSE calculations[24] predict exciton binding energy of approximately 500 meV and 340 meV, respectively, while the experimental observations, from in situ optical absorption and PL spectroscopy, suggest that the corresponding energy is 317 meV.[54] Optical reflectance and spectroscopic data infer the binding energy to be about 412 meV in suspended 1L WSe$_2$.[55] We first summarize our calculated exciton binding energies using the local and long range kernels in Table I and Figure 4. As one can see from Figure 4a, to achieve the experimentally estimated binding energy ~ 400 meV we need $\alpha = 191$ for the contact local kernel, which is an unreasonably high value. It is also clear that the exciton binding energy is very sensitive to the parameter $\alpha$. Notably, the binding energy is very weak if local potential is approximately regarded as "averaged LDA". Similarly, the results for the exciton binding energy depends dramatically on the parameter $\varepsilon$ in the LR potentials, as shown in Figure 4b. The screening parameter $\varepsilon$ is approximately 450 to obtain the experimental value of 400 meV for exciton binding energy. Unfortunately, a realistic screening parameter (~3) cannot produce binding energy that is consistent with the experimental value. Although the calculations of binding energy using local and LR potentials do not include SOC effect, the fact that the obtained results are sensitive to the values of the parameters and one cannot reproduce experimental data with reasonable values of these parameters, suggests that these kernels cannot be used to describe excitons 1L $WSe_2$.

**Table I.** Exciton binding energy (in meV) for 1L $WSe_2$ calculated with the local and LR potentials for different values of corresponding parameters $\alpha$ and $\varepsilon$.

| | | | | | | | | |
|---|---|---|---|---|---|---|---|---|
| Local | $\alpha$ | 1 | 10 | 50 | 100 | 150 | 250 | Exp. 317-412[54,55] |
| | $E_X$ | 0.158 | 1.59 | 37.6 | 163 | 293 | 555 | |
| LR | $\varepsilon$ | 0.05 | 0.01 | 0.0033 | 0.00286 | 0.0025 | 0.002 | |
| | $E_X$ | 0.0589 | 0.2961 | 0.9058 | 5.668 | 202.11 | 493.15 | |

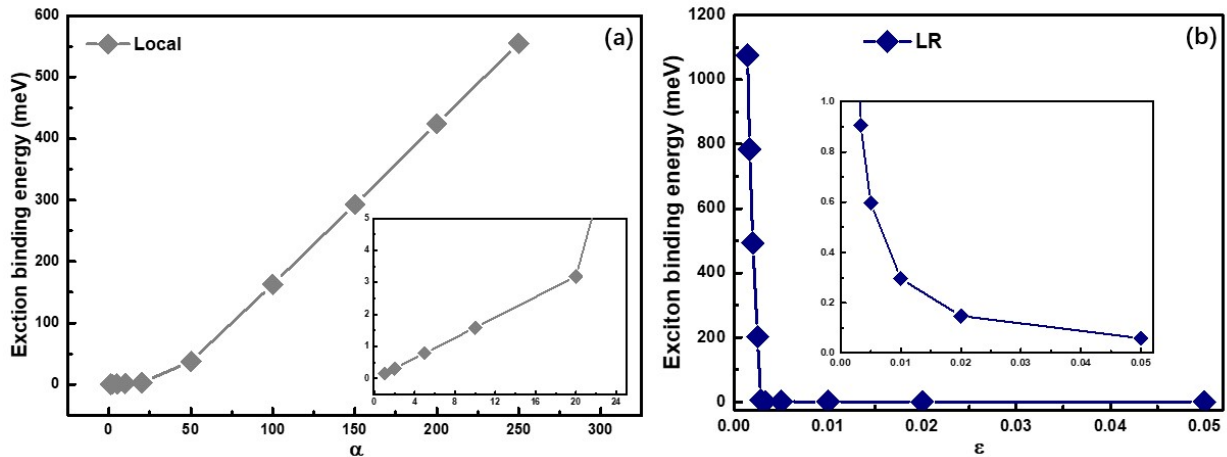

**Figure 4.** (Color online) Exciton binding energy for 1L $WSe_2$ (in meV) calculated with the contact local kernel (a) for different values of coupling $\alpha$ and the LR kernel (b) with different $\varepsilon$.

In order to obtain more physical results for the intra- and intervalley excitons we thus turn to the nanoquanta kernel, which takes into account rather accurately screening in the framework of many-body perturbation theory, and the rest of the computational procedure is parameter-free. In the corresponding kernel calculation, the simple ground states KS potential, such as PBE, can be applied instead of the more complex one in practice. The binding energies of intra- and intervalley excitons calculated with the nanoquanta kernel are listed in the Table II which shows that our values are slightly smaller than those from the experimental data. Some of discrepancy could be attributed to the three-band model that we have used. For bench marking our results, we calculated the binding energy of intravalley bright exciton using the GW-BSE-SOC method. Since this method requires a dense k-mesh and a good number of bands, a $(36 \times 36 \times 1)$ k-point mesh with 10 bands (6 valence bands and 4 conduction bands) were employed. Unfortunately, such a procedure produced a binding energy of only 70 meV for the intravalley bright exciton. Additionally, we notice that the binding energy of bright exciton with SOC differs significantly from that without it. This difference depends mainly on the distinct curvatures of band edges in the SOC and non-SOC cases, which gives rise to differences in the effective masses of carriers.

To shed light on direct and indirect optical transitions, we turn to a comparison of the properties of the intra- and intervalley dark excitons. As shown in Figure 5, *intravalley* dark exciton ($D^0$) is composed of electron and hole with *same* spin, while *intervalley* dark exciton ($I^0$) consists of electron and hole with *opposite* spin. Electron-hole Coulomb interaction is naturally present in both $D^0$ and $I^0$. On the other hand, electron-hole exchange interaction contributes only to the energy of $I^0$ (electron and hole have opposite spins, i.e. electrons from the conduction and valence bands have the same spin) and thus leads to the energy splitting between $D^0$ and $I^0$. We find this splitting using the nanoquanta kernel to be ∼ 22 meV in our calculations, as seen in Table II, in reasonable agreement with experimental observations (9∼16 meV).[17-20] Our calculations of the difference of the exchange parts of the energy of the intra- and inter-valley dark excitons, obtained with the nanoquanta potential, also support this result. Namely, we have calculated the exchange energy related to the inter-valley dark exciton by using an approximate formula, valid for two interacting particles, $-\frac{1}{2}\iint d\vec{r}d\vec{r}'\psi_K^{v*}(\vec{r})\psi_K^{v*}(\vec{r}')f_{XC}(\vec{r}-\vec{r}')\psi_{K'}^{c}(\vec{r}')\psi_{K'}^{c}(\vec{r})$, where $f_{XC}(\vec{r}-\vec{r}')$ is the real-space Fourier transform of the nanoquanta kernel in Eq. (21) (this energy is the analogue of the exchange part of the energy in the Hartree-Fock approximation $-\frac{1}{2}\iint d\vec{r}d\vec{r}'\psi_K^{v*}(\vec{r})\psi_K^{v*}(\vec{r}')\frac{1}{|\vec{r}-\vec{r}'|}\psi_{K'}^{c}(\vec{r}')\psi_{K'}^{c}(\vec{r})$), to find it to be 33meV. Given the simplified two-particle approximation for the excitonic wave function, 33meV is in rather good agreement with the above Casida equation result of 22meV.

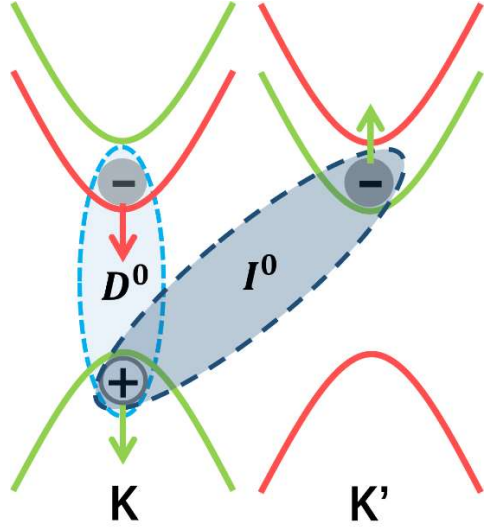

**Figure 5.** (Color online) The energy separation between intravalley dark exciton ($D^0$) and intervalley momentum-forbidden exciton ($I^0$). Spin directions for different bands are shown by arrows.

**Table II**. Binding energy (in meV) of intra- and intervalley excitons with and without SOC for 1L WSe$_2$ calculated with the nanoquanta kernel (Eq. (14) as approximated by Eq. (15)-(18)).

|  | Exciton | Binding energy |
|---|---|---|
|  | Intravalley dark ($D^0$) | 132 |
| SOC | Intervalley dark ($I^0$) | 110 |
|  | Intravalley bright | 91 |
| No SOC | Intravalley bright | 154 |

3.2 Transition contribution maps

For a deeper understanding of the properties of the dark excitons, we calculated momentum-resolved weights of the electron-hole transitions for each excitonic state. This was done by calculating the eigenvectors from Eq. (14) and plotting the transition contribution map - square of the modulus of the components of eigenvectors (that correspond to different momenta). The obtained transition contribution maps (Figure 6), that show contribution of different "DFT-state excitations" to the excitonic transitions, facilitate a better understanding of excitonic states. In this map, in the xy-plane we show the momentum of the electron excited state, while the hole excited state is assumed to be at the K-point. Our findings reveal that electron-hole transitions consist dominantly of states nearby the K and K′ valleys in the case of intravalley excitons (both intravalley bright and dark excitations), while the excitations from the adjacent area of K and K′ also contribute to the optical transition. For the intervalley electron-hole excitations, the primary contributions of excitonic hybridizations are from the areas around K and K′ valleys. Surprisingly, there are additional contributions from the Γ point.

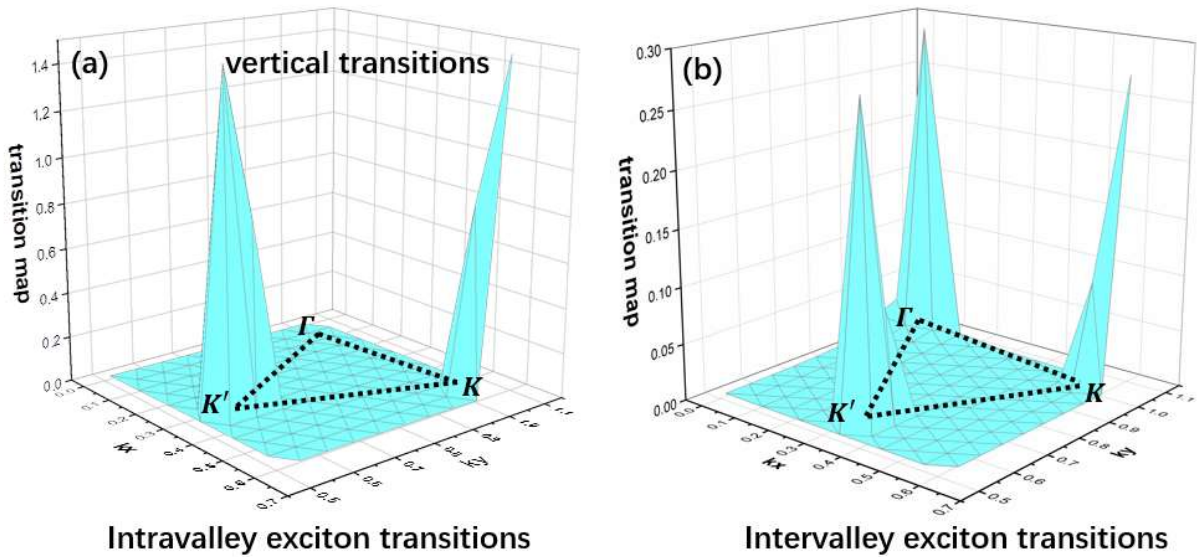

**Figure 6.** (Color online) The transition contribution map for the intra- (a) and intervalley (b) excitons. The final momenta of electrons are given in the xy-plane and the final momenta of holes are located at K point.

3.3 Energy shifts and lifetimes due to charge-phonon interaction

Inclusion of electron-phonon interactions, as summarized in Section 2.1, allows calculation of the temperature dependencies of the excitation energy by documenting the shifts in the position of the inter-band and exciton transition peaks. In particular, by using Eq. (26) we find that the bandgap shrinks from 1.46 eV to 1.396 eV (decrease by 64 meV) as temperature increases from 0K to 300 K (Figure 7c). The top of valence band lifts and bottom of conduction band drops simultaneously in this temperature range. The corresponding variation of ~100meV in electron and hole energies at 300 K are well in agreement with the experimental observations.[56] In this work, we did not calculate change of the energy of different types of excitons with temperature, which is a separate computationally challenging project. We analyzed the phonon part of lifetime of the electron, hole (from the inverse of the imaginary part of self-energy (24)) and bright exciton (equal to inverse of the corresponding line broadening (27)). We found that this quantity also changes significantly with temperature (Figures 7d and 7e). These results point to the need to include phonon effects to describe experimental spectra of the system.

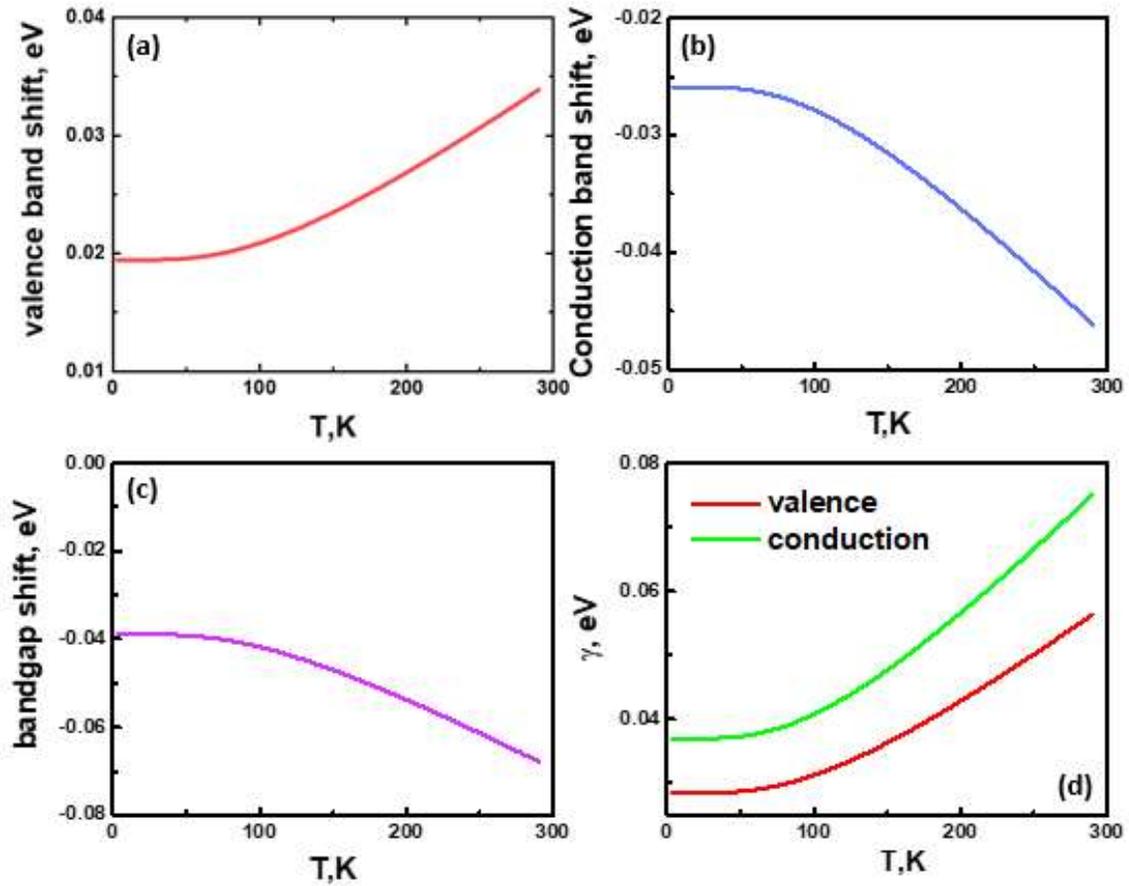

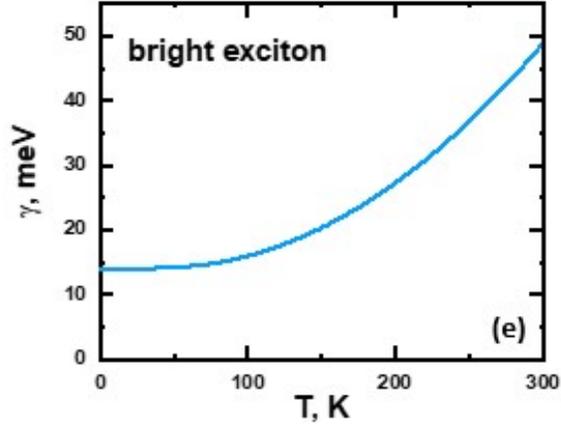

**Figure 7.** (Color online) The energy renormalization as a function of temperature (0~300 K) in the valence band (a), conduction band (b), and the bandgap (c). The inverse lifetime of particles (holes in the valence band and electrons in the conduction band) (d) and of the bright exciton (e) as a function of temperature.

## 4. Conclusions

In this work, we have formulated a density-matrix TDDFT approach to examine the properties of intervalley excitons in 1L WSe$_2$. We applied different XC kernels to analyze the excitonic properties of the system, and found that the nanoquanta kernel provides physically meaningful results. This TDDFT approach is computationally much more feasible as compared to the BSE method. The exchange energy-splitting of the energies of intra- and intervalley dark excitons obtained by the nanoquanta kernel is approximately 22 meV, which is in good agreement with the experimental results. The TDDFT calculations of the transition contribution map for the excitations in 1L WSe$_2$ show that the electron-hole transitions near the K and K′ valleys mostly contribute to the formation of intravalley excitonic states. Surprisingly, in the inter-valley exciton case states around Γ-point also give a significant contribution.

We have also applied second-order perturbation theory to evaluate the contribution of electron-phonon scattering to the temperature-induced shift of the bands and of the bandgap, as well as to calculate the electron-, hole- and bright exciton lifetimes. We found that these quantities depend significantly on temperature, and that our results for the bandgap shift are in agreement with available experimental data. These results suggest that phonons cannot be neglected in analysis of the optical spectra of the system. The formalism presented here is ready to be applied to describe systems with many (orbital, spin, valley, lattice vibration) degrees of freedom in TMDs and other materials.

**Acknowledgement**

This work was supported in part by DOE grant DE-FG02-07ER46354 and by a grant of the China Scholarship Council (J.S). Computations were carried out at High Performance Computing Facility STOKES at UCF.18